\documentclass[twocolumn,pre,twoside,floatfix]{revtex4-1}

\usepackage{amsmath}
\usepackage{amssymb}
\usepackage{graphicx}
\usepackage{dcolumn}
\usepackage{bm}
\usepackage{color}
\usepackage{multirow}
\usepackage{hyperref}
\usepackage{epstopdf}

\usepackage[utf8]{inputenc}
\usepackage[T1]{fontenc}
\usepackage{microtype}
\usepackage{txfonts}

\begin{document}


\title{Electrostatic interaction between dissimilar colloids at fluid interfaces}

\author{Arghya Majee}
\email{majee@is.mpg.de}
\author{Timo Schmetzer}
\author{Markus Bier}
\email{bier@is.mpg.de}
\affiliation{Max-Planck-Institut f\"ur Intelligente Systeme, Heisenbergstr.\ 3, 70569 Stuttgart, Germany}
\affiliation{IV. Institut f\"{u}r Theoretische Physik, Universit\"{a}t Stuttgart, Pfaffenwaldring 57, 70569 Stuttgart, Germany}


\begin{abstract}
The electrostatic interaction between two \textit{non-identical}, moderately charged colloids situated in 
close proximity of each other at a fluid interface is studied. By resorting to a well-justified model system, 
this problem is analytically solved within the framework of linearized Poisson-Boltzmann (PB) density
functional theory (DFT). The resulting interaction comprises a surface and a line 
part, both of which, as functions of the inter-particle separation, show a rich behavior including monotonic as 
well as non-monotonic variations. In almost all cases, these variations cannot be captured correctly by using 
the superposition approximation. Moreover, expressions for the surface tensions, the line tensions and
the fluid-fluid interfacial tension, which are all independent of the inter-particle separation, are obtained.
Our results are expected to be particularly useful for emulsions  stabilized by oppositely 
charged particles.
\end{abstract}

\maketitle


\section{Introduction}

Interfaces between two immiscible fluids like oil and water are often energetically very expensive, which
is quantified by a high interfacial tension. Therefore, systems featuring fluid interfaces try to reduce 
their free energy by minimizing the interfacial area. When colloids are suspended in either of the fluid phases, 
such a reduction is easily achieved via entrapment of the particles at the interface \cite{Ram03}: a well-known
phenomenon that forms the basis of any Pickering emulsion \cite{Pic07}. This finds application in diverse fields 
including biomedicine, cosmetic industry, oil recovery, water purification, anti-reflective coating and so 
on \cite{Bin06, Ray09, Isa17}. One major advantage of particle stabilized emulsions is that the use of risky
surfactants can be avoided, which is particularly important for cosmetic, food and pharmaceutical
applications \cite{Abe98, Mao11, Mao12}. In the form of colloidosome \cite{Din02}, it can even be used to 
prepare inorganic protocells \cite{Li13}.

Emulsions can be stabilized by using a single type of particle or by using a mixture of particles that differ 
in their charge. Over the last two decades, this latter has been shown to be useful at many instances \cite{Abe98, 
Abe01, Bin08, Mao11, Mao12, Nal14, Nal15}. For example, intrinsically hydrophilic particles that adsorb weakly 
to interfaces, can form flocs of increased hydrophobicity when mixed with oppositely charged particles and 
attach stably to an interface \cite{Bin08}. Sometimes relatively strongly charged particles, when approaching 
the interface from the aqueous phase, alone cannot attach to oil-water interfaces due to strong repulsive 
interaction with the image charge formed in presence of the dielectric jump at the interface \cite{Wan12}. 
Aggregates of oppositely charged particles facilitate adsorption in this case owing to a reduction of the 
net charge of the aggregate compared to the single particle \cite{Nal14, Nal15}. Moreover, even for chemically 
identical interacting particles, specially at short separations, identical charge densities at the two 
particles are not always guranteed \cite{Maj17}.

Due to the steep trapping potential (typically several orders of magnitude larger than the thermal energy $k_BT$) 
felt by the particles at an interface, their out-of-plane movement is essentially frozen \cite{Bin06}. On the 
other hand, the in-plane movement and the structure of the resulting monolayer depends sensitively on the 
inter-particle interactions. Depending upon the system, particle interaction may consist of different contributions
like van der Waals, capillary, electrostatic, steric or magnetic interactions. Here we focus on charged colloids and 
consider exclusively the electrostatic interaction between them. At low particle concentration, this can be 
described by a simple dipolar interaction where each particle and its asymmetric counter-ion cloud 
generate a dipole perpendicular to the interface \cite{Pie80, Hur85}. However, for systems with oppositely charged 
particles, where the separation distance can be well below a nanometer, or for dense suspensions such a 
simple picture cannot be applied.   

A semi-analytical solution for interaction between nonuniformly but \textit{like-charged} particles at an 
air-water interface has been recently provided which is valid at all separations between the particles \cite{Lia16}. 
An alternate approach, which makes the problem completely analytically solvable, is to use a relatively simple 
yet reasonable model system by ignoring the particle curvature and treating them as flat plates due to 
their short separation \cite{Maj14, Maj16}. Using this model system with two plates, effective electrostatic 
pair-interaction between \textit{identical} colloids situated in the close vicinity of each other has been 
studied analytically both within and beyond the superposition approximation \cite{Maj14}.
However, the interaction of two identical colloidal particles is a situation which is typically not met
in reality, e.g., due to various kinds of polydispersities generated during the preparation process, and a 
proper theory describing the interaction between \textit{non-identical} particles situated close to each other 
at a fluid interface is not available so far.

\begin{figure}[!t]
\centering{\includegraphics[width=8.4cm]{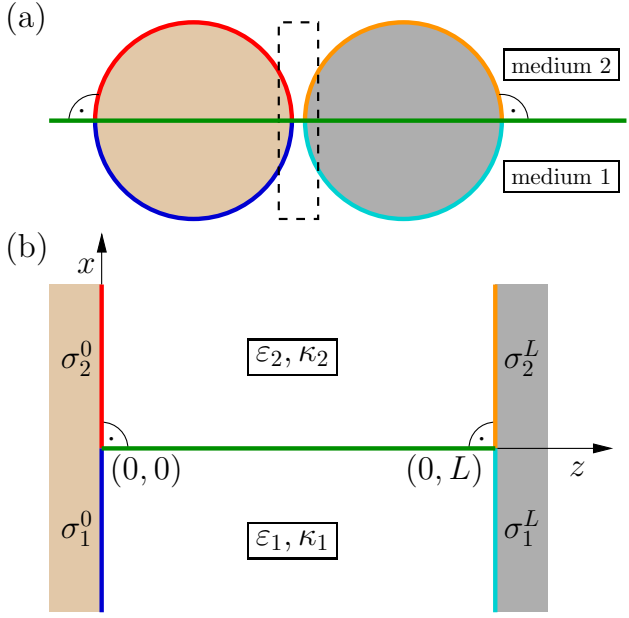}}
\caption{(a) Sketch of the system under consideration. A pair of non-identical spherical colloids trapped at an oil-water 
             interface (indicated by the horizontal green line) and separated from each other by a distance small compared to 
             their radii. We consider neutral-wetting situations corresponding to $90^\circ$ liquid-particle 
             contact angles for all particles. (b) Sketch of the simplified system that models the boxed region in panel (a). Due 
             to their short separation, particles are approximated as flat walls positioned at $z=0$ and $L$. The region in 
             between is filled with two immiscible fluids forming an interface at $x=0$. Permittivity and the inverse Debye 
             length of fluid ``1'' (``2'') are given by $\varepsilon_1$ ($\varepsilon_2$) and $\kappa_1$ ($\kappa_2$), 
             respectively. The surfaces carry fixed charge densities ($\sigma_1^0$, $\sigma_1^L$, $\sigma_2^0$, and $\sigma_2^L$) 
             which, in general, may not be the same for the two walls and can vary over each wall depending upon the 
             fluid phase it is in contact with.}
\label{Fig1}
\end{figure}

Therefore, here we address this problem by using the aforementioned model system of flat plates with a
liquid-liquid interface in between; see Fig.~\ref{Fig1}. Additionally, 
we discard any deformation of the fluid-fluid interface and consider it to be planar with a $90^\circ$ liquid-particle 
contact angle. This implies that the particles are dipped equally deep into both fluid phases 
such that the corresponding reduction of the interfacial area is maximal. Similar situations have been 
encountered in previous experimental studies \cite{Ave03, Rei04, Mas10, Pet14, Pet16, Kra16} and have been 
used in theoretical modelling \cite{Nal14, Upp14, Pus15, Pet16, Kra16, Bos16} of such systems as well. 
There is strong experimental evidence that charges can be present at the particle-oil interface 
\cite{Ave00, Ave02, Par08, Gao14, Kel15}. Accordingly, particles, or the plates of our model system, are assumed 
to be charged in both the fluid phases.
The most salient feature of our system is now the presence of potentially disparate but uniform charge densities
on each of the four colloid-fluid contact surfaces.

In this work the model system, as depicted in Fig.~\ref{Fig1}(b), is analyzed within the framework of classical
DFT which allows one to derive the governing equations for the electrostatic potential as well as the boundary 
conditions and to calculate the resulting effective interactions in a self-consistent manner. First, the electrostatic 
potential distribution inside the system is calculated once by avoiding and once by using the superposition approximation. 
In what follows, results obtained in the former case will be denoted as exact results. Next, the interaction between 
the charged solid surfaces, which can be decomposed into contributions proportional to the areas of the solid surfaces 
and the lengths of the three-phase contact lines, is calculated using both the exact and the superposition 
electrostatic potentials. Finally, a comparison of the exact and superposition results is presented for a wide 
range of system parameters. In addition, our results also provide insight into a particularly striking observation 
regarding identical particles: all the effective interaction energies of the exact calculation 
were reported to differ by a factor of 2 from those obtained within the superposition approximation in the asymptotic 
limit \cite{Maj14}. A conceivable explanation includes the symmetry of the system or the total number of colloidal
particles. As will be shown below, the present analysis clearly rules out the former possibility.

\section{Formalism}

We confine ourselves to a space filled with two immiscible fluids and bounded by two parallel charged walls placed at
$z=0$ and $z=L$ in a Cartesian coordinate system; see Fig.~\ref{Fig1}(b). The fluid-fluid interface is perpendicular to
the walls and situated at $x=0$. 
The fluid phase spanning the lower (upper) half-space corresponding to $x<0$ ($x>0$) is denoted 
as medium ``1'' (``2''). The added salt is a binary compound with monovalent components only and the bulk ionic strengths
in medium $i\in\{1,2\}$ are $I_i$. Since the walls are not identical, surface charge density $\sigma(\mathbf{r})$, in 
general, is a function of both the spatial coordinates. 
At contact with medium $i\in\left\{1,2\right\}$, the charge densities at the left ($z=0$) and at the right 
($z=L$) walls are given by $\sigma_i^0$ and $\sigma_i^L$, respectively. 
In general, salt ions, charged surfaces and the fluid-fluid interface influence the structure of the fluids 
which in turn leads to the number densities $n_{\pm}(\mathbf{r})$ of the positive and negative ions. 
However, these structures form on the scale of the bulk correlation length, 
which usually falls much below the length scale to be considered here. As a result, the fluids are modeled as structureless 
continuous linear dielectric media characterized by dielectric constants $\varepsilon_i=\varepsilon_{r,i}\varepsilon_0$, 
$i\in\left\{1,2\right\}$, where $\varepsilon_{r,i}$ is the relative permittivity of medium $i$ and $\varepsilon_0$ is 
the permittivity of vacuum. 
Therefore, on sufficiently large length scales, the overall fluid dielectric profile $\varepsilon(\mathbf{r})$ varies
step-like at the interface: $\varepsilon(x<0)=\varepsilon_1$ and $\varepsilon(x>0)=\varepsilon_2$. The charge density 
$e(n_{+}(\mathbf{r})-n_{-}(\mathbf{r}))$ in each medium, however, varies on the scale of the respective Debye lengths
$\kappa_i^{-1}=\sqrt{\varepsilon_{r,i}/\left(8\pi\ell_BI_i\right)}$ which set the length scales of our interest and are 
much larger than the bulk correlation length \cite{Bie12}. Here $\ell_B=e^2/\left(4\pi\varepsilon_0k_BT\right)$
is the vacuum Bjerrum length with the elementary charge $e>0$, the Boltzmann constant $k_B$ and the temperature $T$.

Following the standard description at a mean-field level, we treat the ions as point-like objects and ignore their correlation. 
Considering the bulks of both fluid media as reservoirs with which ions are exchanged the grand 
canonical density functional corresponding to our system in units of the thermal energy $k_BT=1/\beta$ is then given by
\begin{align}
   \beta\Omega\left[n_{\pm}\right] = \int\limits_Vd^3r\Bigg[
   &\sum\limits_{k=\pm}n_k\left(\mathbf{r}\right)
    \Bigg\{\ln\left(\frac{n_k\left(\mathbf{r}\right)}{\zeta_k}\right)-1
    +\beta V_k\left(\mathbf{r}\right)\Bigg\}\notag\\
   &+\frac{\beta\mathbf{D}
    \left(\mathbf{r},\left[n_\pm\right]\right)^2}{2\varepsilon\left(\mathbf{r}\right)}\Bigg],
\label{eq:1}
\end{align}
with the fugacities $\zeta_{\pm}$ of the two ion-species, the electric displacement $\mathbf{D}$, and the solvation
free energies $V_{\pm}(\mathbf{r})$ of the ions. The integration volume $V$ is the slab region ($0\leq z\leq L$) 
enclosed by the two walls. The electric displacement $\mathbf{D}$ fulfills Gauss' law 
\begin{align}
   \nabla\cdot\mathbf{D}\left(\mathbf{r},\left[n_{\pm}\right]\right)=e\left(n_+(\mathbf{r})-n_-(\mathbf{r})\right)
\label{eq:1b}
\end{align}
in the interior of $V$ whereas it is fixed by the given surface charge densities at the walls. The first two terms in 
the first line of Eq.~(\ref{eq:1}) is the entropic ideal gas contribution of the ions whereas the last term within the 
curly brackets describes the contribution due to ion-solvent interaction. The last term represents the electric field 
energy due to the ion distribution and the surface charge densities. First, $\widetilde\Omega\left[n_{\pm}\right]$ is 
obtained from $\Omega\left[n_{\pm}\right]$ by expanding the latter in terms of small deviations of the ion number 
densities from the bulk ionic strengths and retaining terms up to quadratic order (for details see Chapter~II of 
Ref.~\cite{Sch17}). Subsequently $\widetilde\Omega\left[n_{\pm}\right]$ is minimized with respect to the ionic density 
profiles $n_{\pm}$ to obtain the equilibrium profiles 
\begin{align}
   n_{\pm}^{\text{eq}}(\mathbf{r})=I_i\left(1\mp\beta e\left\{\Psi_i(\mathbf{r})-\Psi_{b,i}\right\}\right)
\label{eq:1a}
\end{align}
in the two media. Noting the relation $\mathbf{D}=-\varepsilon_i\nabla\Psi_i$ between the electric displacement $\mathbf{D}$ 
and the electrostatic potential $\Psi_i(\mathbf{r})$ in medium $i\in\{1,2\}$, Eqs.~(\ref{eq:1b}) and (\ref{eq:1a}) 
ultimately lead to the linearized PB or Debye-H\"uckel (DH) equation
\begin{align}
 \Delta\Psi_i(\mathbf{r})=\kappa_i^2\left(\Psi_i(\mathbf{r})-\Psi_{b,i}\right),
\label{eq:2}
\end{align}
subject to the following 
boundary conditions: (i) the electrostatic potential must remain finite in the limit $x\rightarrow\pm\infty$, (ii) both the 
electrostatic potential and the $x$-component of the electric displacement vector should be continuous at the fluid-fluid
interface, i.e., $\Psi_1(x=0^-)=\Psi_2(x=0^+)$ and $\varepsilon_1\partial_x\Psi_1(x=0^-)=\varepsilon_2\partial_x\Psi_2(x=0^+)$,
and (iii) the $z$-component of the displacement vector should match the charge densities at the surfaces: 
$\varepsilon_i\partial_z\Psi_i(z=0)=-\sigma_i^0$ and $\varepsilon_i\partial_z\Psi_i(z=L)=\sigma_i^L$.
These boundary conditions imply the global charge neutrality for our system. $\Psi_{b,i}$ in Eqs.~(\ref{eq:1a}) and (\ref{eq:2}) 
represents the electrostatic potential in the bulk of the medium $i\in\{1,2\}$. It is constructed such that $\Psi_{b,1}=0$ 
and $\Psi_{b,2}=\Psi_D$ where $\Psi_D$ originates from an unequal partitioning of ions close to the interface in
between the two media and it is known as the Donnan potential or Galvani potential difference between the two fluid
phases \cite{Bag06}. If the solvation free energies of the ions are given by $V_{\pm}(\mathbf{r})=0$ in medium ``1'' and 
$V_{\pm}(\mathbf{r})=f_{\pm}$ in medium ``2'', then the Donnan potential can be written as \cite{Maj14, Sch17}
\begin{align}
 \Psi_D=-\frac{1}{2e}\left(f_+-f_-\right).
\label{eq:2a}
\end{align}

In order to calculate the equilibrium grand potential $\widetilde\Omega(L)$ of the system with wall separation $L$ we first
rewrite $\widetilde\Omega\left[n_{\pm}\right]$ as a functional of the electrostatic 
potential $\Psi$ by inserting the equilibrium ion density profiles $n_{\pm}^{\text{eq}}[\Psi]$ considered as
functionals of $\Psi$. Then, by plugging the expressions for $\Psi_i(x,z)$ obtained by solving Eq.~(\ref{eq:2}), 
one obtains $\widetilde\Omega(L)=\widetilde\Omega\left[n_{\pm}[\Psi]\right]$ which is composed of the following different 
contributions:
\begin{align}
 \widetilde\Omega(L)= &\sum\limits_{i\in\{1,2\}}\left[\Omega_{b,i}V_i+\left(\frac{\gamma_i^0+\gamma_i^L}{2}+
                   \omega_{\gamma,i}(L)\right)A_i\right]\notag\\
                  &+\gamma_{1,2}A_{1,2}+\left(\frac{\tau^0+\tau^L}{2}+\omega_{\tau}(L)\right)\ell,
\label{eq:3}
\end{align}
where $\Omega_{b,i}$ is bulk grand potential per volume of medium $i\in\{1,2\}$; $V_i$ is the volume of medium $i$;
$\gamma_i^0$ and $\gamma_i^L$ are the surface tensions acting between medium ``$i$'' and the 
wall present at $z=0$ and $z=L$, respectively; $\omega_{\gamma,i}(L)$ is the effective interaction energy between 
surface elements in contact with and acting through medium $i$, $A_i$ is the total area of the two surfaces in contact with medium $i$,
$\gamma_{1,2}$ is the interfacial tension acting between medium $1$ and $2$; $A_{1,2}$ is the total area of the fluid-fluid
interface; $\tau^0$ and $\tau^L$ are the line tensions acting at the two three-phase contact lines formed at $z=0$ and $z=L$,
respectively; and $\omega_{\tau}(L)$ is the effective interaction energy between the contact lines expressed per total 
length $\ell$ of the two contact lines. Here $\gamma_i^0$, $\gamma_i^L$, $\tau^0$, and $\tau^L$ describe the 
interaction of a single wall with its surrounding fluid(s) and are thus $L$-independent. Therefore, the only 
$L$-dependent quantities in Eq.~(\ref{eq:3}) are the surface interaction energies $\omega_{\gamma,i}(L)$ and the line 
interaction energy $\omega_{\tau}(L)$. Please note that in the limit of infinitely large separation between the walls 
these interaction contributions vanish: $\omega_{\gamma,i}(L\rightarrow\infty)\rightarrow0$ and $\omega_{\tau}(L\rightarrow\infty)\rightarrow0$.

The ultimate goal here is to infer properties of the total effective interaction between two colloidal particles in 
Fig.~\ref{Fig1}(a).
As the total volume of medium $i\in\{1,2\}$ (see the space outside of the colloids in Fig.~\ref{Fig1}(a)) and the total area
of the fluid-fluid interface (see the green line outside of the colloids in Fig.~\ref{Fig1}(a)) do not change as functions
of the distance between the colloids, i.e., of $L$.
Hence the terms involving $\Omega_{b,i}$ and $\gamma_{1,2}$ in Eq.~(\ref{eq:3}), which quantifies only the contributions from
inside the dashed box in Fig.~\ref{Fig1}(a), are compensated by contributions from outside the box when one considers the 
effective interaction of the total system and can be disregarded in the following. Therefore, the total effective inter-surface 
interaction is solely determined by the two quantities $\omega_{\gamma,i}(L)$ and $\omega_{\tau}(L)$.

\section{Results and discussion}

\subsection{Electrostatic potential}

\subsubsection{Exact calculation}

The \textit{e}xact electrostatic potential $\Psi_i^e(x,z)$ everywhere inside our system must satisfy the DH 
equation (Eq.~(\ref{eq:2})) along with the boundary conditions listed below Eq.~(\ref{eq:2}). In order to achieve 
such a solution, we first split the actual problem depicted in Fig.~\ref{Fig1}(b) into three sub-problems: (i) two walls 
situated at $z=0$ and $z=L$ with charge densities $\sigma_1^0$ and $\sigma_1^L$, respectively, and the space in 
between the walls is filled with medium ``1'' only, (ii) the same set-up as in part (i) but with the fluid medium 
``2'' instead of ``1'' and consequently, the charge densities at the walls replaced by $\sigma_2^0$ and $\sigma_2^L$, 
and (iii) two fluids separated by an interface at $x=0$ in the absence of any walls. For all these sub-problems,
the electrostatic potential is obtained by solving the DH equation. Since this is a linear equation, we, by fiat, 
add the solution of sub-problem (i) with the solution obtained for the lower half space ($x<0$) in sub-problem (iii) 
and the solution of sub-problem (ii) with the one obtained for the upper half-space ($x>0$) in sub-problem (iii). 
It turns out that the resulting solutions after performing such additions satisfy all the required boundary conditions 
except for the continuity of the electrostatic potential at the interface. This discrepancy is eliminated by adding 
a correction function which is also solution of the DH equation and which takes care of the continuity problem at the 
interface while keeping the already satisfied boundary conditions unaltered. Construction of a function like this 
is accomplished by means of Fourier series expansion (for details, please refer to Chapter~III of Ref.~\cite{Sch17}). 
Finally, putting everything together, one arrives at the following final expressions for the electrostatic potential 
in the two media:
\begin{widetext}
 \begin{align}
  \Psi_i^e(x,z)=&\frac{\sigma_i^0\cosh\left(\kappa_i\left(L-z\right)\right)+\sigma_i^L\cosh\left(\kappa_iz\right)}
    {\varepsilon_i\kappa_i\sinh\left(\kappa_iL\right)}
    +\Psi_{b,i}
    +\sum\limits_{j\in\{1,2\}}^{j\neq i}\Bigg[
    \frac{(-1)^j\kappa_j\varepsilon_j\Psi_D}{\kappa_1\varepsilon_1+\kappa_2\varepsilon_2}e^{-\kappa_i\lvert x\rvert}
    +\frac{1}{\kappa_1\varepsilon_1+\kappa_2\varepsilon_2}\left(\frac{\sigma_j^0+\sigma_j^L}{\kappa_j}-
    \frac{\kappa_j\varepsilon_j}{\kappa_i\varepsilon_i}\frac{\sigma_i^0+\sigma_i^L}{\kappa_i}\right)
    \frac{e^{-\kappa_i\lvert x\rvert}}{L}\notag\\
  & +\sum\limits_{n=1}^{\infty}\frac{2}{p_1^{(n)}\varepsilon_1+p_2^{(n)}\varepsilon_2}
    \left(\frac{\sigma_j^0+(-1)^n\sigma_j^L}{p_j^{(n)}}
    -\frac{p_j^{(n)}\varepsilon_j}{p_i^{(n)}\varepsilon_i}\frac{\sigma_i^0+(-1)^n\sigma_i^L}{p_i^{(n)}}\right)
    \frac{e^{-p_i^{(n)}\lvert x\rvert}}{L}\cos\left(\frac{n\pi z}{L}\right)\Bigg],
  \label{eq:4}
 \end{align}
\end{widetext}
with $p_m^{(n)}=\sqrt{\left(\frac{n\pi}{L}\right)^2+\kappa_m^2}$. The first term of this expression represents 
the solution of the sub-problem (i) or (ii) depending upon the medium where the potential is looked at, the 
second term and the first term of the sum over $j$ together is the solution of sub-problem (iii) and the 
rest is the correction function. Some of the limiting cases and boundary conditions can be verified easily from 
Eq.~(\ref{eq:4}). For example, in the limit $x\rightarrow\pm\infty$, all the terms which vary exponentially with $x$  
vanish and one is left with the first two terms which are the solutions inside the two media in the presence of 
the walls but far away from the fluid-fluid interface at $x=0$. In the limit $L\rightarrow\infty$ with $\sigma_1^L=\sigma_2^L=0$, 
the first term of Eq.~(\ref{eq:4}) simplifies to an exponentially decaying potential of a single charged wall, 
the summation term $\propto1/L$ in the first line vanishes, and the sum over $n$ in the second line can be 
converted to an integral over $q=n\pi/L$; see Eq.~(\ref{eq:5}) below. 
Therefore, Eq.~(\ref{eq:4}) reduces to the potential for a single charged 
wall placed at $z=0$ in contact with two electrolytes spanning the region $z>0$. 
For both $x,L\rightarrow\infty$, only the first two terms do not vanish with the first term reduced to an exponentially 
decaying function characteristic of a single charged wall in contact with an electrolyte. 
Finally, again from Eq.~(\ref{eq:4}), at $z=0$ and $z=L$, 
one easily verifies the relations $\varepsilon_i\partial_z\Psi_i^e(z=0)=-\sigma_i^0$ and $
\varepsilon_i\partial_z\Psi_i^e(z=L)=\sigma_i^L$ since the $z$-derivative of all the terms except the first 
one vanish. When all the charge densities are set to zero, only the second and the third 
terms in Eq.~(\ref{eq:4}) do not vanish which forms the solution of the sub-problem (iii) and, on top of that, if one 
approaches $x\rightarrow\pm\infty$, $\Psi_i^e$ reduces to $\Psi_{b,i}$ which is the potential in the bulk of two 
media far away from the interface.

\subsubsection{Superposition approximation}

Since DH equation is linear, a superposition approximation, in principle, can be used to obtain the potential 
distribution for a two-body problem. This requires one to first calculate the potential due to each object in the 
absence of the other and then to simply add these two potentials at each point in space to obtain an approximate
potential in the presence of both particles. Accordingly, we first calculate the electrostatic potential due to
a single charged wall placed at $z=0$ and carrying charge densities $\sigma_i^0$ in contact with medium $i\in\{1,2\}$
filling the space $z>0$ with the fluid-fluid interface formed at $x=0$. Following the same procedure as described 
above, we again divide the problem into the following sub-problems: (i) a single wall placed at $z=0$ in contact 
with medium ``1'' and carrying charge density $\sigma_1^0$, (ii) a single wall place at $z=0$ with medium ``2'' 
and carrying charge density $\sigma_2^0$, (iii) two fluid media separated by an interface at $x=0$ in the 
absence of any wall. By adding the solutions of these problems in the respective medium, one obtains a solution 
which satisfies all the required boundary conditions except the continuity of the potential at the interface. This 
is rectified by constructing again a correction function in the same fashion as before albeit by means of Fourier 
transforms. Finally, the electrostatic potential due to a \textit{sin}gle charged wall at $z=0$ is given by
\begin{align}
 &\Psi_i^{\text{sin}}(x,z)=\frac{\sigma_i^0}{\varepsilon_i\kappa_i}e^{-\kappa_iz}
 +\Psi_{b,i}+\sum\limits_{j\in\{1,2\}}^{j\neq i}\Bigg[
 \frac{(-1)^j\kappa_j\varepsilon_j\Psi_D}{\kappa_1\varepsilon_1+\kappa_2\varepsilon_2}e^{-\kappa_i\lvert x\rvert}\notag\\
 & +\frac{1}{\pi}\int\limits_{-\infty}^{\infty}dq
 \frac{e^{-p_i(q)\lvert x\rvert}\cos(qz)}{p_1(q)\varepsilon_1+p_2(q)\varepsilon_2}
 \left(\frac{\sigma_j^0}{p_j(q)}-\frac{p_j(q)\varepsilon_j}{p_i(q)\varepsilon_i}\frac{\sigma_i^0}{p_i(q)}\right)\Bigg],
 \label{eq:5}
\end{align}
with $p_m(q)=\sqrt{q^2+\kappa_m^2}$. From this expression, one readily verifies the relation 
$\varepsilon_i\partial_z\Psi_i^{\text{sin}}(z=0)=-\sigma_i^0$. In the limit $x\rightarrow\pm\infty$ all the terms that 
vary exponentially with $x$ vanish and one is left with the first two terms in Eq.~(\ref{eq:5}) which is nothing but the 
usual screened potential due to a single charged wall plus the potential $\Psi_{b,i}$ due to partitioning of ions at the 
interface. Setting all the charge densities to zero, one recovers the potential distribution due to the presence of a fluid 
interface in the absence of any wall and on top of that, in the limit $x\rightarrow\pm\infty$, Eq.~(\ref{eq:5}) 
reduces to the bulk potential $\Psi_{b,i}$.

Next, a similar solution is obtained for a wall placed at $z=L$ carrying charge densities $\sigma_1^L$ and $\sigma_2^L$ 
in the absence of any wall at $z=0$. It can be obtained readily from Eq.~(\ref{eq:5}) without an explicit calculation 
by just replacing $z$ and $\sigma_i^0$ with $L-z$ and $\sigma_i^L$, respectively. Then by adding these two 
solutions for walls present at $z=0$ and $z=L$ in the absence of each other, one obtains the following expression 
for the electrostatic potential $\Psi_i^s(x,z)$ in the presence of both walls under the \textit{s}uperposition approximation:
\begin{widetext}
 \begin{align}
  \Psi_i^s(x,z)&=\frac{\sigma_i^0e^{-\kappa_iz}+\sigma_i^Le^{\kappa_i(z-L)}}{\kappa_i\varepsilon_i}
  +2\Psi_{b,i}+\sum\limits_{j\in\{1,2\}}^{j\neq i}
  \frac{(-1)^j2\kappa_j\varepsilon_j\Psi_D}{\kappa_1\varepsilon_1+\kappa_2\varepsilon_2}e^{-\kappa_i\lvert x\rvert}\notag\\
  & +\sum\limits_{j\in\{1,2\}}^{j\neq i}\frac{1}{\pi}\int\limits_{-\infty}^{\infty}dq
  \frac{(-1)^je^{-p_i(q)\lvert x\rvert}}{p_1(q)\varepsilon_1+p_2(q)\varepsilon_2}
  \left\{\left(\frac{\sigma_j^0}{p_j(q)}-\frac{p_j(q)\varepsilon_j}{p_i(q)\varepsilon_i}\frac{\sigma_i^0}{p_i(q)}\right)\cos(qz)
  +\left(\frac{\sigma_j^L}{p_j(q)}-\frac{p_j(q)\varepsilon_j}{p_i(q)\varepsilon_i}\frac{\sigma_i^L}{p_i(q)}\right)
  \cos\left(-q(z-L)\right)\right\}.
  \label{eq:6}
 \end{align}
\end{widetext}
It is worth noting that $\Psi_i^s(x,z)$ fails to behave properly in most of the limiting cases and to satisfy the 
boundary conditions listed below Eq.~(\ref{eq:2}). At the two boundaries, $\varepsilon_i\partial_z\Psi_i^s(z=0)\neq-\sigma_i^0$ 
and $\varepsilon_i\partial_z\Psi_i^s(z=L)\neq\sigma_i^L$. In the limit $L\rightarrow\infty$ with $\sigma_1^L=\sigma_2^L=0$, 
$\Psi_i^s(x,z)$ does not 
reduce to the expression for the potential due to a single wall at $z=0$ given in Eq.~(\ref{eq:5}). Setting all the 
charge densities to zero and $x\rightarrow\pm\infty$, one does not even recover the bulk potential $\Psi_{b,i}$. We like 
to finish our discussion on the electrostatic potential by pointing out that for the case of identical particles, i.e., 
for $\sigma_i^0=\sigma_i^L$, Eqs.~(\ref{eq:4}) and (\ref{eq:6}) reduce to Eq.~(1) and (2) of Ref.~\cite{Maj14}, respectively. 
Please note that the walls in Ref.~\cite{Maj14} are situated at $z=\pm L$ whereas in the current set-up they are situated at 
$z=0$ and $L$ which needs to be taken into account for proper transformations. Plots showing comparisons of the two 
potential distributions (Eqs.~(\ref{eq:4}) and (\ref{eq:6})) are displayed in Chapter III of Ref.~\cite{Sch17}.

\begin{figure*}[!t]
\centering{\includegraphics[width=15cm]{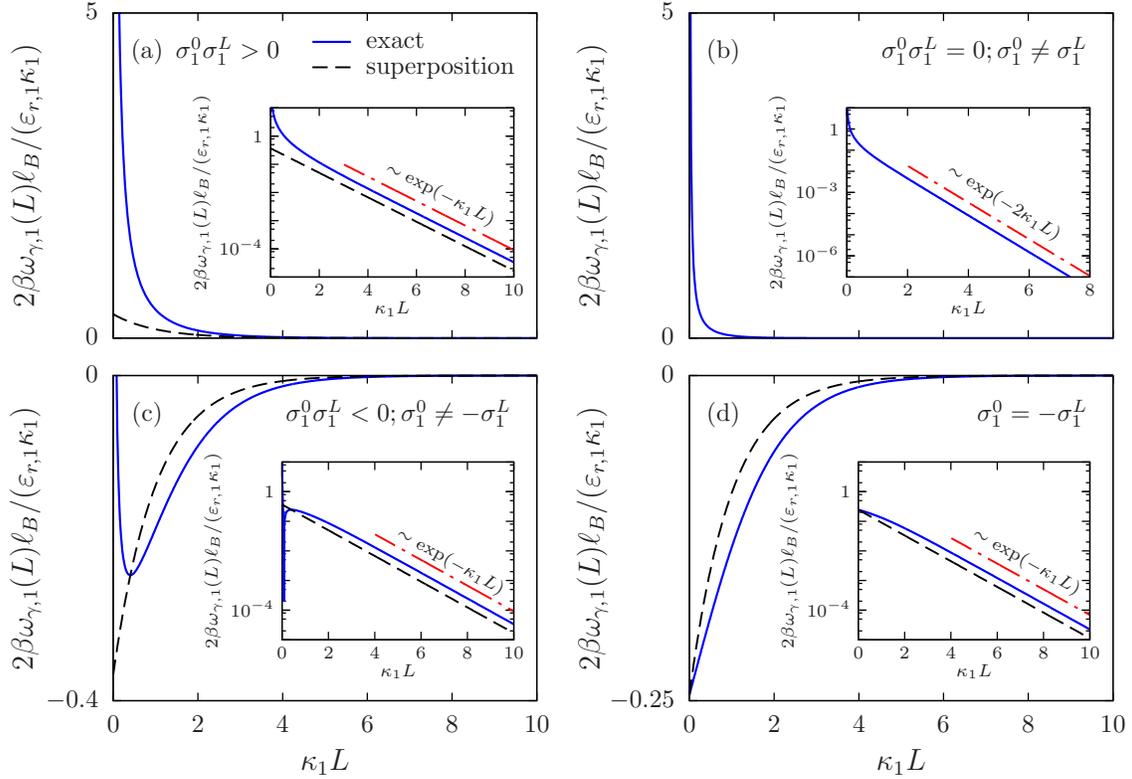}}
\caption{Variation of the surface interaction energy densities $\omega_{\gamma,1}^e(L)$ and $\omega_{\gamma,1}^s(L)$, 
         both expressed in units of $\varepsilon_{r,1}\kappa_1/\left(2\beta\ell_B\right)$, as functions of the scaled
         separation $\kappa_1L$ between the walls for (a) $\sigma_1^0\sigma_1^L>0$, (b) $\sigma_1^0\sigma_1^L=0$
         with $\sigma_1^0\neq\sigma_1^L$, (c) $\sigma_1^0\sigma_1^L<0$ with $\sigma_1^0\neq-\sigma_1^L$, and 
         (d) $\sigma_1^0=-\sigma_1^L$. As shown by the plots, $\omega_{\gamma,1}^e(L)$ diverges in the limit 
         of vanishing separation between the walls except for the case considered in panel (d), whereas 
         $\omega_{\gamma,1}^s(L)$ is always finite in this limit. For the special case of $\sigma_1^0=-\sigma_1^L$
         considered in panel (d), both $\omega_{\gamma,1}^e(0)$ and $\omega_{\gamma,1}^s(0)$ are finite and 
         $\omega_{\gamma,1}^e(0)=\omega_{\gamma,1}^s(0)$. In the opposite limit, i.e., in the limit of infinitely 
         large separations between the walls, both $\omega_{\gamma,1}^e$ and $\omega_{\gamma,1}^s$, if non-zero, 
         decay exponentially which in each case is confirmed by the semi-logarithmic plots in the insets. 
         However, when one of the two charge densities are zero (panel (b)), $\omega_{\gamma,1}^s(L)$ 
         is zero for any separation between the walls whereas $\omega_{\gamma,1}^e(L)$ is non-zero but decays 
         twice as slow compared to the other cases. The overall decay of the surface interaction 
         energies are monotonic except when the walls are oppositely charged with $\sigma_1^0\neq-\sigma_1^L$ 
         In this case, the exact surface interaction energy shows a minimum before decaying to zero at large separations.}
\label{Fig2}
\end{figure*}

\subsection{Interaction parameters}

Once the electrostatic potential is known, it can be inserted back into the grand potential functional to
obtain $\widetilde\Omega(L)=\widetilde\Omega\left[n_{\pm}[\Psi]\right]$ and all the interaction parameters described 
in Eq.~(\ref{eq:3}) can be extracted. This is done 
by identifying the terms proportional to $V_i$, $A_i$, $A_{1,2}$ and $\ell$, and separating the $L$-independent parts from the 
$L$-dependent parts in the expression for $\widetilde\Omega(L)$. Calculating interaction parameters within the superposition 
approximation implies using expression for the electrostatic potential given in Eq.~(\ref{eq:6}) but no further superposition 
is performed for any other quantities.

\subsubsection{Surface interaction energy}

Surface interaction energy density $\omega_{\gamma,i}(L)$, as defined in Eq.~(\ref{eq:3}), is given within the \textit{e}xact 
calculation by
\begin{align}
 \omega_{\gamma,i}^e(L)=\left(\left(\sigma_i^0\right)^2\!+\!\left(\sigma_i^L\right)^2\right)
                        \frac{\coth\left(\kappa_iL\right)\!-\!1}{4\kappa_i\varepsilon_i}
                        \!+\!\frac{\sigma_i^0\sigma_i^L}{2\kappa_i\varepsilon_i\sinh\left(\kappa_iL\right)}.
 \label{eq:7}
\end{align}
For $\sigma_i^0=\sigma_i^L=0$, there is no electrostatic interaction between the surfaces, i.e., $\omega_{\gamma,i}^e(L)=0$ for 
any separation $L$ between the walls. 
In case of $\sigma_i^0=-\sigma_i^L$, using L'H\^opital's 
rule, one obtains from Eq.~(\ref{eq:7}) a finite, i.e., non-divergent, attractive surface interaction in
the limit of vanishing separations: $\omega_{\gamma,i}^e(L\rightarrow0)=-\frac{\left(\sigma_i^0\right)^2}{2\kappa_i\varepsilon_i}$. 
In contrast, for $\sigma_i^0\neq-\sigma_i^L$, both $\coth(\kappa_iL)$ and $\frac{1}{\sinh(\kappa_iL)}$ diverge $\sim\frac{1}{L}$, 
and, as a result, $\omega_{\gamma,i}^e$ diverges to $+\infty$ in the limit $L\rightarrow0$, i.e., $\omega_{\gamma,i}^e(L\to0)$ 
becomes repulsive. In the opposite limit, i.e., for $L\rightarrow\infty$, one also has two different cases depending upon the 
charge densities. For $\sigma_i^0\sigma_i^L\neq0$, $\omega_{\gamma,i}^e(L\rightarrow\infty)\simeq\frac{1}{\kappa_i\varepsilon_i}\left[
\frac{\left(\sigma_i^0\right)^2+\left(\sigma_i^L\right)^2}{2}e^{-2\kappa_iL}+\sigma_i^0\sigma_i^Le^{-\kappa_iL}\right]$, 
i.e., to the leading order $\omega_{\gamma,i}^e$ varies $\sim e^{-\kappa_iL}$. However, if one of the two charge densities 
$\sigma_i^0$ and $\sigma_i^0$ vanishes, the decay becomes twice as slow $\left(\sim e^{-2\kappa_iL}\right)$. 
The latter effect is related to the finite size of the system. The overall decay of $\omega_{\gamma,i}^e(L)$ is monotonic unless 
the conditions $\sigma_i^0\sigma_i^L<0$ and $\sigma_i^0\neq-\sigma_i^L$ are simultaneously satisfied. In this case, 
$\omega_{\gamma,i}^e(L\rightarrow0)\rightarrow+\infty$ and one can verify easily from Eq.~(\ref{eq:7}) that 
$\omega_{\gamma,i}^e(L)$ has a zero at 
$L_0=-\frac{1}{\kappa_i}\ln\left(\frac{-2\sigma_i^0\sigma_i^L}{\left(\sigma_i^0\right)^2+\left(\sigma_i^L\right)^2}\right)$, 
and an extremum at $L_{\text{ex}}=\frac{1}{\kappa_i}\operatorname{arcosh}\left(-\frac{1}{2}\left(\frac{\sigma_i^0}{\sigma_i^L}
+\frac{\sigma_i^L}{\sigma_i^0}\right)\right)$. Since $\omega_{\gamma,i}^e(L\rightarrow0)\rightarrow+\infty$, this extremum 
must be a minimum. Therefore, with increasing separation $L$, the surface interaction energy density is initially positive, 
then becomes negative, shows a minimum and eventually vanishes for large separations as $\sim e^{-\kappa_iL}$.

Within the \textit{s}uperposition approximation, the surface interaction energy density is given by
\begin{align}
 \omega_{\gamma,i}^s(L)=\frac{\sigma_i^0\sigma_i^L}{2\kappa_i\varepsilon_i}e^{-\kappa_iL}.
 \label{eq:8}
\end{align}
In the special case of $\sigma_i^0=-\sigma_i^L$, $\omega_{\gamma,i}^s(0)=\omega_{\gamma,i}^e(0)$. However, for 
$\sigma_i^0\sigma_i^L=0$, there is no surface interaction within the superposition approximation, i.e., 
$\omega_{\gamma,i}^s(L)=0$ for any given separation between the walls which is not the case for the exact solution where one 
has a non-zero surface interaction even if one of the two surface charge densities vanishes. 
Moreover, for $\sigma_i^0\sigma_i^L\neq0$, 
contrary to the results obtained within the exact calculation, $\omega_{\gamma,i}^s(L)$ is always (i) finite in the limit of 
vanishing separation between the walls, and (ii) decays monotonically with the separation length $L$. Although Eq.~(\ref{eq:8}) 
predicts the exponential decay at large separations correctly, the prefactor is too small by a factor of 2 compared to the exact 
result, i.e., $\omega_{\gamma,i}^e(L\rightarrow\infty)/\omega_{\gamma,i}^s(L\rightarrow\infty)=2$. Please note that a similar 
factor of 2 is present for the case of identically charged walls as well \cite{Maj14} which, as is confirmed by the present 
analysis, cannot be related to a symmetry of the surface charge distribution. 

The surface interaction energy densities (expressed in units of 
$\varepsilon_{r,1}\kappa_1/\left(2\beta\ell_B\right)$) in medium ``1'' within the exact (Eq.~(\ref{eq:7})) and the 
superposition (Eq.~(\ref{eq:8})) approach as functions of the scaled separation $\kappa_1L$ between 
the walls are compared in Fig.~\ref{Fig2} for different combinations of the charge densities $\sigma_1^0$ and $\sigma_1^L$ at the 
walls. Please note that for typical system parameters the scale-factor $\varepsilon_{r,1}\kappa_1/\left(2\beta\ell_B\right)$ 
in aqueous electrolyte solution with $\kappa_1=0.1\,\mathrm{nm}^{-1}$ is $\approx0.3\,\mathrm{mN/m}$. Although not shown 
in the plots, $\omega_{\gamma,2}^e(L)$ and $\omega_{\gamma,2}^s(L)$ behave similarly to $\omega_{\gamma,1}^e(L)$ 
and $\omega_{\gamma,1}^s(L)$, respectively. 

\begin{figure*}[!t]
\centering{\includegraphics[width=16cm]{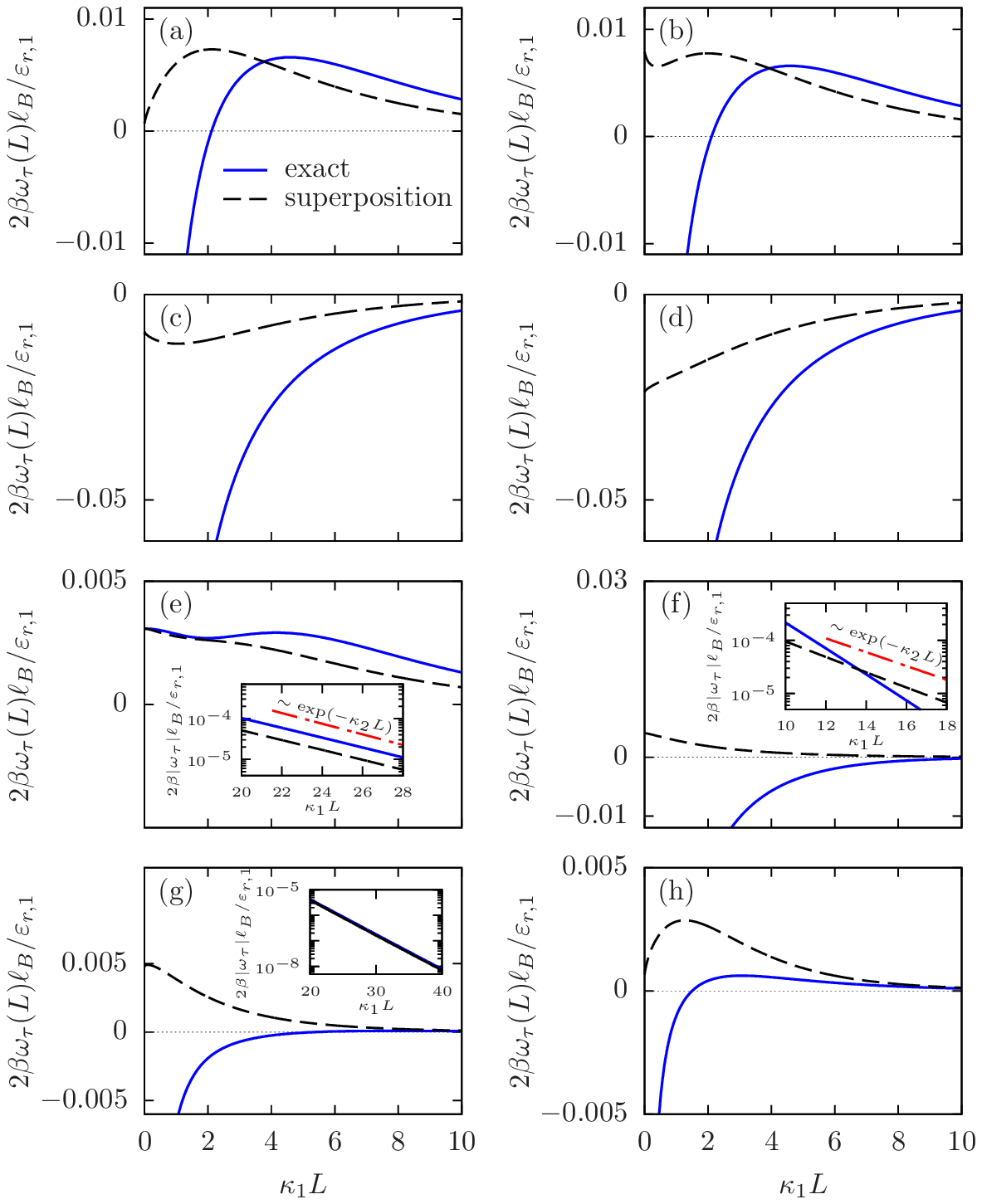}}
\caption{Line interaction energy $\omega_{\tau}(L)$ within both the exact and the superposition calculations expressed in  
         units of $\varepsilon_{r,1}/\left(2\beta\ell_B\right)$ as functions of the scaled separation length $\kappa_1L$ 
         for different combinations of the charge densities $\sigma_1^0$, $\sigma_1^L$, $\sigma_2^0$, and $\sigma_2^L$
         specified in Table~\ref{Tab1}. As it 
         is clear from the plots, both $\omega_{\tau}^e$ and $\omega_{\tau}^s$ can vary monotonically as well as non-monotonically 
         depending upon the parameters and in most cases the superposition approximation fails to capture the 
         correct behavior properly. In the limit of vanishing separation between the walls, $\omega_{\tau}^s$ stays finite, 
         whereas $\omega_{\tau}^e$ usually diverges unless the two walls are \textit{exactly} oppositely charged 
         (panel (e)). The decay at large separations is always exponential (shown, for example, in the inset of panel (e) 
         but the decay rate for $\omega_{\tau}^e$ and $\omega_{\tau}^s$ differs by a factor of 2 when one of the walls 
         is uncharged as shown by the semi-logarithmic plot in the inset of panel (f)). Moreover, even when the superposition
         expression predicts the correct decay rate it always underestimates the magnitude by a factor of 2 if none of the four 
         charge densities are zero. This can be seen from the corresponding curves in the inset of panel (e), which, instead 
         of falling on top of each other, are parallel to each other. However, when either one of the four 
         or two diagonally opposite charge densities are zero, the ratio 
         $\omega_{\tau}^e(L\rightarrow\infty)/\omega_{\tau}^s(L\rightarrow\infty)$ converges to a value which depends 
         upon the system parameters. For the parameters used in panel (g) and (h), it is close to unity.}
\label{Fig3}
\end{figure*}

\begin{table*}
\caption{Values used for the plots in Fig.~\ref{Fig3} of the charge densities $\sigma_1^0$, $\sigma_1^L$, $\sigma_2^0$, and 
         $\sigma_2^L$ in units of 
         $e/\mathrm{nm}^2$, inverse Debye lengths $\kappa_1$ and $\kappa_2$ in medium ``1'' and ``2'', respectively in 
         units of $\mathrm{nm}^{-1}$, relative permittivities $\varepsilon_{r,1}$ and $\varepsilon_{r,2}$ of 
         medium ``1'' and ``2'', respectively, the Donnan potential $\Psi_D$ in units of $1/(\beta e)$ and the Bjerrum 
         length $\ell_B$ in units of $\mathrm{nm}$.}
\renewcommand{\arraystretch}{1.2} 
\centering
\resizebox{1.0\textwidth}{!}
{
\begin{tabular}{cllllllllll}
\hline
\hline
Figure~3 & \multicolumn{1}{c}{$\sigma_1^0 (e/\mathrm{nm}^2)$} & \multicolumn{1}{c}{$\sigma_1^L (e/\mathrm{nm}^2)$} 
& \multicolumn{1}{c}{$\sigma_2^0 (e/\mathrm{nm}^2)$} & \multicolumn{1}{c}{$\sigma_2^L (e/\mathrm{nm}^2)$} 
& \multicolumn{1}{c}{$\kappa_1 (\mathrm{nm}^{-1})$} & \multicolumn{1}{c}{$\kappa_2 (\mathrm{nm}^{-1})$} 
& \multicolumn{1}{c}{$\varepsilon_{r,1}$} & \multicolumn{1}{c}{$\varepsilon_{r,2}$} 
& \multicolumn{1}{c}{$\beta e\Psi_D$} & \multicolumn{1}{c}{$\ell_B (\mathrm{nm})$} \\ 
\hline
(a) & $\phantom{-}0.02$ & $\phantom{-}0.03$ & $-0.0004$ & $\phantom{-}0.0002$ & $0.1$ & $0.03$ & $80$ & $2$ & $1$ & $55.7$ \\

(b) & $-0.02$ & $-0.03$ & $\phantom{-}0.0004$ & $-0.0002$ & $0.1$ & $0.03$ & $80$ & $2$ & $1$ & $55.7$ \\

(c) & $\phantom{-}0.02$ & $-0.03$ & $\phantom{-}0.0004$ & $\phantom{-}0.0002$ & $0.1$ & $0.03$ & $80$ & $2$ & $1$ & $55.7$ \\

(d) & $-0.02$ & $\phantom{-}0.03$ & $-0.0004$ & $-0.0002$ & $0.1$ & $0.03$ & $80$ & $2$ & $1$ & $55.7$ \\

(e) & $\phantom{-}0.02$ & $-0.02$ & $\phantom{-}0.0002$ & $-0.0002$ & $0.1$ & $0.03$ & $80$ & $2$ & $1$ & $55.7$ \\

(f) & $\phantom{-}0.02$ & $\phantom{-}0$ & $\phantom{-}0.0004$ & $\phantom{-}0$ & $0.1$ & $0.03$ & $80$ & $2$ & $1$ & $55.7$ \\

(g) & $\phantom{-}0.02$ & $\phantom{-}0$ & $\phantom{-}0$ & $\phantom{-}0.0002$ & $0.1$ & $0.03$ & $80$ & $2$ & $1$ & $55.7$ \\

(h) & $\phantom{-}0.02$ & $\phantom{-}0.03$ & $\phantom{-}0$ & $\phantom{-}0.0002$ & $0.1$ & $0.03$ & $80$ & $2$ & $1$ & $55.7$ \\
\hline
\hline
\end{tabular}
}
\label{Tab1}
\end{table*}

\begin{figure*}[!t]
\centering{\includegraphics[width=17.2cm]{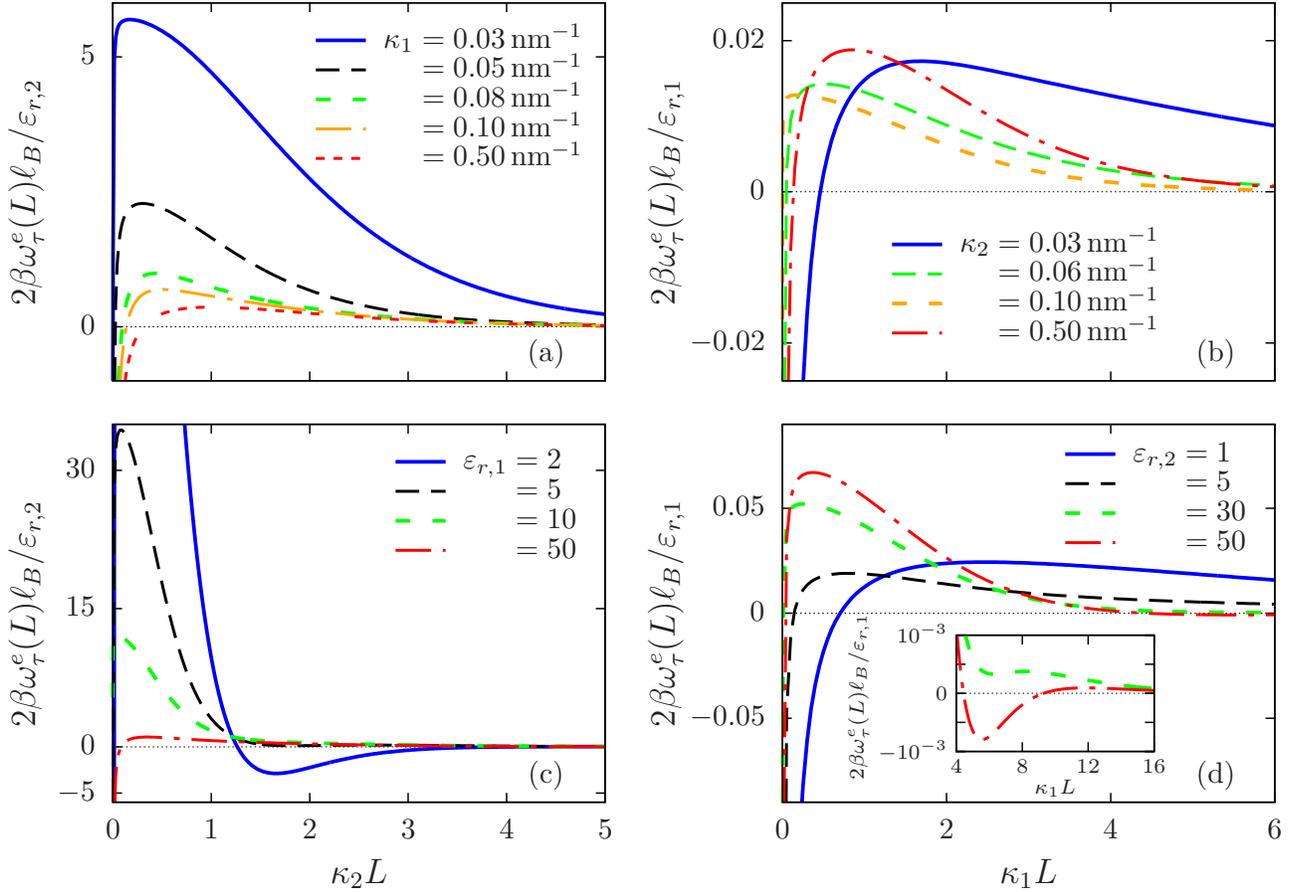}}
\caption{Variation of the exact line interaction energy density $\omega_{\tau}^e(L)$ (expressed in units of 
         $\varepsilon_{r,2}/\left(2\beta\ell_B\right)$ in the left panels and $\varepsilon_{r,1}/\left(2\beta\ell_B\right)$ 
         in the right panels) as function of the scaled separation $\kappa_2L$ (left panels) and $\kappa_1L$ (right 
         panels) for varying (a) inverse Debye length $\kappa_1$ in medium ``1'', (b) inverse Debye length $\kappa_2$ in medium ``2'', 
         (c) relative permittivity $\varepsilon_{r,1}$ of medium ``1'', and (d) relative permittivity $\varepsilon_{r,2}$ of 
         medium ``2''. The other parameters used for the plots are: $\sigma_1^0=0.02\,\mathrm{e/nm}^2$, 
         $\sigma_1^L=-0.03\,\mathrm{e/nm}^2$, $\sigma_2^0=-0.0004\,\mathrm{e/nm}^2$, $\sigma_2^L=0.0002\,\mathrm{e/nm}^2$, 
         and $\beta e\Psi_D=1$. Unless otherwise stated, $\kappa_1=0.1\,\mathrm{nm}^{-1}$, $\kappa_2=0.03\,\mathrm{nm}^{-1}$, 
         $\varepsilon_{r,1}=80$, and $\varepsilon_{r,2}=2$ are considered. As one can see, $\omega_{\tau}^e(L)$ always show
         at least one maximum. However, with decreasing $\varepsilon_{r,1}$, a single minimum can occur while with increasing 
         $\varepsilon_{r,2}$, a minimum and a second maximum can occur.}
\label{Fig4}
\end{figure*}

\begin{figure}[!t]
\centering{\includegraphics[width=8.5cm]{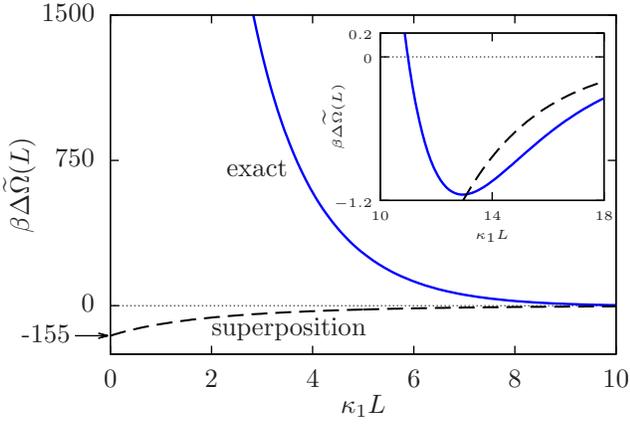}}
\caption{Total interaction energy $\Delta\widetilde\Omega(L)$ ($=\omega_{\gamma,1}(L)A_1+\omega_{\gamma,2}(L)A_2
         +\omega_{\tau}(L)\ell$) between the two surfaces in the units of $1/\beta$ as a function of the scaled separation $\kappa_1L$ for a system with 
         $\sigma_1^0=0.001\,\mathrm{e/nm}^2$, $\sigma_1^L=-0.1\,\mathrm{e/nm}^2$, $\sigma_2^0=0.0001\,\mathrm{e/nm}^2$,
         $\sigma_2^L=-0.01\,\mathrm{e/nm}^2$, $\kappa_1=0.1\,\mathrm{nm}^{-1}$, $\kappa_2=0.03\,\mathrm{nm}^{-1}$, $\varepsilon_{r,1}=80$,
         $\varepsilon_{r,2}=2$, $\beta e\Psi_D=1$, $\ell_B=55.7\,\mathrm{nm}$, $A_1\approx7600\,\mathrm{nm}$, $A_2\approx22000\,\mathrm{nm}$, and 
         $\ell\approx240\,\mathrm{nm}$. The total effective surface areas $A_i$ and the total effective length $\ell$ of the three-phase
         contact lines are calculated as explained in the main text for particles of radii $R=100\,\mathrm{nm}$.
         As one can see, whereas the superposition approximation (black dashed line) predicts a deep minimum at vanishing separation between 
         the surfaces leading to aggregation, the exact result (blue solid line) is mostly positive with a shallow minimum (see the inset) 
         at around $\kappa_1L\approx13$ and thus, rules out the possibility of aggregation.}
\label{Fig5}
\end{figure}

\subsubsection{Line interaction energy}

The line interaction energy density $\omega_{\tau}(L)$, as defined in Eq.~(\ref{eq:3}), is given within the \textit{e}xact calculation by
\begin{widetext}
 \begin{align}
  \omega_{\tau}^e(L)&=\frac{1}{L}\frac{1}{\kappa_1\varepsilon_1+\kappa_2\varepsilon_2}
  \left(\frac{\left(\sigma_1^0+\sigma_1^L\right)\left(\sigma_2^0+\sigma_2^L\right)}{2\kappa_1\kappa_2}
  -\frac{\kappa_2\varepsilon_2}{\kappa_1\varepsilon_1}\left(\frac{\sigma_1^0+\sigma_1^L}{2\kappa_1}\right)^2
  -\frac{\kappa_1\varepsilon_1}{\kappa_2\varepsilon_2}\left(\frac{\sigma_2^0+\sigma_2^L}{2\kappa_2}\right)^2\right)\notag\\
  &+\frac{1}{2L}\sum\limits_{n=1}^{\infty}\frac{1}{p_1^{(n)}\varepsilon_1+p_2^{(n)}\varepsilon_2}
  \left(\frac{2\left(\sigma_1^0+(-1)^n\sigma_1^L\right)\left(\sigma_2^0+(-1)^n\sigma_2^L\right)}{p_1^{(n)}p_2^{(n)}}
  -\frac{p_2^{(n)}\varepsilon_2}{p_1^{(n)}\varepsilon_1}\left(\frac{\sigma_1^0+(-1)^n\sigma_1^L}{p_1^{(n)}}\right)^2
  -\frac{p_1^{(n)}\varepsilon_1}{p_2^{(n)}\varepsilon_2}\left(\frac{\sigma_2^0+(-1)^n\sigma_2^L}{p_2^{(n)}}\right)^2\right)\notag\\
  &-\frac{1}{2\pi}\int\limits_0^{\infty}\frac{dq}{p_1(q)\varepsilon_1+p_2(q)\varepsilon_2}
  \left(\frac{2\left(\sigma_1^0\sigma_2^0+\sigma_1^L\sigma_2^L\right)}{p_1(q)p_2(q)}
  -\frac{p_2(q)\varepsilon_2}{p_1(q)\varepsilon_1}\left(\frac{\left(\sigma_1^0\right)^2+\left(\sigma_1^L\right)^2}{p_1(q)^2}\right)
  -\frac{p_1(q)\varepsilon_1}{p_2(q)\varepsilon_2}\left(\frac{\left(\sigma_2^0\right)^2+\left(\sigma_2^L\right)^2}{p_2(q)^2}\right)\right),
  \label{eq:9}
 \end{align}
and within the \textit{s}uperposition approximation by
 \begin{align}
  \omega_{\tau}^s(L)&=\frac{\kappa_1\varepsilon_1\Psi_D}{4\left(\kappa_1\varepsilon_1+\kappa_2\varepsilon_2\right)}
  \left(\frac{\sigma_2^0+\sigma_2^L}{\kappa_2}-\frac{\kappa_2\varepsilon_2}{\kappa_1\varepsilon_1}\frac{\sigma_1^0+\sigma_1^L}{\kappa_1}\right)
  -\frac{\Psi_D}{4\pi}\int\limits_{-\infty}^{\infty}\frac{p_1(q)\varepsilon_1}{p_1(q)\varepsilon_1+p_2(q)\varepsilon_2}
  \left(\frac{\sigma_2^0+\sigma_2^L}{p_2(q)}-\frac{p_2(q)\varepsilon_2}{p_1(q)\varepsilon_1}\frac{\sigma_1^0+\sigma_1^L}{p_1(q)}\right)
  \frac{\sin\left(qL\right)}{q}dq\notag\\
  &+\frac{1}{2\pi}\int\limits_{-\infty}^{\infty}\frac{\cos(qL)dq}{p_1(q)\varepsilon_1+p_2(q)\varepsilon_2}
  \left(\frac{\sigma_1^0\sigma_2^L+\sigma_2^0\sigma_1^L}{p_1(q)p_2(q)}
  -\frac{p_2(q)\varepsilon_2}{p_1(q)\varepsilon_1}\frac{\sigma_1^0\sigma_1^L}{p_1(q)^2}
  -\frac{p_1(q)\varepsilon_1}{p_2(q)\varepsilon_2}\frac{\sigma_2^0\sigma_2^L}{p_2(q)^2}\right),
  \label{eq:10}
 \end{align}
\end{widetext}
with $p_m^{(n)}$ and $p_m(q)$ as defined earlier after Eq.~(\ref{eq:4}) and Eq.~(\ref{eq:5}), respectively. In general, in 
the limit of vanishing separation between the walls, $\omega_{\tau}^e$ in Eq.~(\ref{eq:9}) diverges $\sim\frac{1}{L}$. On the 
other hand, in the limit $L\rightarrow0$, the second term of Eq.~(\ref{eq:10}) vanish whereas the other terms have a finite 
non-zero value. Consequently, $\omega_{\tau}^s$ also remains finite in this limit. In the opposite limit, i.e., for 
$L\rightarrow\infty$, the sum in the second line of Eq.~(\ref{eq:9}) is cancelled by the integral in the third line. The 
first term also vanishes in this limit. However, the difference of the sum and the integral in Eq.~(\ref{eq:9}) such that 
$\omega_{\tau}^e$ decays exponentially to zero in the limit of large separation between the walls. The superposition 
expression (\ref{eq:10}) also decays exponentially to zero in the limit of large separation. As it is difficult to analyze 
the overall variation of the line interactions analytically, we have plotted it in Fig.~\ref{Fig3} for different 
combination of the four charge densities $\sigma_1^0$, $\sigma_1^L$, $\sigma_2^0$, and $\sigma_2^L$, which are specified in
Table~\ref{Tab1}. As one can see from the plots, unless the 
two walls are \textit{exactly} oppositely charged (as it is the case in Fig.~\ref{Fig3}(e)), the line interaction within 
the exact calculation diverges whereas it remains finite for the superposition approximation in the limit $L\rightarrow0$. 
For exactly oppositely charged walls, both $\omega_{\tau}^e$ and $\omega_{\tau}^s$ remain finite and have the same value at 
$L=0$ (see Fig.~\ref{Fig3}(e)). If all four charge densities are non-zero (cases considered in 
Figs.~\ref{Fig3}(a)-\ref{Fig3}(e); see Table \ref{Tab1}), the exponential decay at large separations 
is characterized by $\sim e^{-\kappa_2L}$ since we have assumed $\kappa_2<\kappa_1$ (see Table~\ref{Tab1}). The decay is shown 
only in Fig.~\ref{Fig3}(e) for the sake of neatness. However, even when both $\omega_{\tau}^e$ and $\omega_{\tau}^s$ decay 
as $\sim e^{-\kappa_2L}$, the latter is too small compared to the former by a factor of 2 (see for example, the inset of 
Fig.~\ref{Fig3}(e)). If one of the walls is uncharged, $\omega_{\tau}^s$ still varies as $\sim e^{-\kappa_2L}$ but 
$\omega_{\tau}^e$ decays as $\sim e^{-2\kappa_2L}$ (see Fig.~\ref{Fig3}(f)). Not only that, if two diagonally opposite 
charge densities out of the four are zero or if only one of the four charge densities is zero (cases considered in 
Figs.~\ref{Fig3}(g) and \ref{Fig3}(h); see Table \ref{Tab1}), or if two of the four charge densities, which are facing 
each other, are zero, then the ratio $\omega_{\tau}^e(L\rightarrow\infty)/\omega_{\tau}^s(L\rightarrow\infty)$ converges 
to a value which is not generally fixed but which depends upon the parameters of the system, especially the Donnan potential 
$\Psi_D$ (for details, see section 16.3.10 of Ref.~\cite{Sch17}). As is clear from Fig.~\ref{Fig3}, the overall variation 
of both, $\omega_{\tau}^e(L)$ and $\omega_{\tau}^s(L)$, with the separation distance $L$ can be non-monotonic but in most cases 
the superposition result 
cannot capture the correct variation. Another important observation is that the exact expression for the line interaction 
energy density (Eq.~(\ref{eq:9})) is insensitive to changes in the signs of the charge densities when the sign of all four 
of them are reversed. Clearly, the superposition expression (Eq.~(\ref{eq:10})) lacks this property in general as 
Eq.~(\ref{eq:10}) contains $\sigma$-terms which are odd as well as $\sigma$-terms which are even. This particular feature can 
also be verified by comparing Figs.~\ref{Fig3}(a) and \ref{Fig3}(b) or Figs.~\ref{Fig3}(c) and \ref{Fig3}(d). We have 
checked the other combinations of the four charge densities as well, but concluded that the behavior can be broadly 
classified into the eight different cases presented in Fig.~\ref{Fig3}.

Figure~\ref{Fig4} shows the variation of the line interaction $\omega_{\tau}^e(L)$ within the exact calculation as function 
of the separation length between the walls for different values of the inverse Debye lengths and permittivities of the two 
liquids. For this plot we consider $\sigma_1^0$ and $\sigma_2^L$ to be positive and $\sigma_1^L$ and $\sigma_2^0$ to be 
negative, a case that is not considered in Fig.~\ref{Fig3}. As one can see, $\omega_{\tau}^e(L)$ always shows at least a maximum, 
the position of which shifts to larger separations with increasing inverse Debye length $\kappa_1$ of medium ``1'' or decreasing 
inverse Debye length $\kappa_2$ of medium ``2''. With decreasing relative permittivity $\varepsilon_{r,1}$ of medium ``1'', 
$\omega_{\tau}^e(L)$ can show a single minimum whereas with increasing relative permittivity $\varepsilon_{r,2}$ of medium ``2'', 
it can show a minimum and an additional maximum. Please note that while varying the quantities of medium ``1'', i.e. in the 
left panels of Fig.~\ref{Fig4}, the horizontal axis is scaled with $\kappa_2=0.03\,\mathrm{nm}^{-1}$. Therefore, a maximum 
even at $\kappa_2L\approx0.3$ corresponds to $L\approx10\,\mathrm{nm}$, which is well above the molecular dimension.

It is worth mentioning that the corresponding expressions for the identical particles given in Ref.~\cite{Maj14} can 
be recovered from Eqs.~(\ref{eq:7})--(\ref{eq:10}) by setting $\sigma_1^0=\sigma_1^L$, $\sigma_2^0=\sigma_2^L$, and $L=2L'$ 
where $L'$ is the separation between the identically charged walls. This last transformation is needed as the walls in 
Ref.~\cite{Maj14} were considered to be separated by a distance of $2L$ instead of $L$.

At this point it is natural to ask whether the differences between the exact and the superposition calculations are significant 
and whether the line part of the interaction can play any significant role compared to the surface contributions for real experimental 
setups. To answer the first question, we plot in Fig.~\ref{Fig5} the total interaction energy 
$\Delta\widetilde\Omega(L)=\omega_{\gamma,1}(L)A_1+\omega_{\gamma,2}(L)A_2+\omega_{\tau}(L)\ell$ between two surfaces in the units of $1/\beta$ 
as a function of the separation distance $\kappa_1L$ for a typical experimental system with $\sigma_1^0=0.001\,\mathrm{e/nm}^2$, $\sigma_1^L=-0.1\,\mathrm{e/nm}^2$, 
$\sigma_2^0=0.0001\,\mathrm{e/nm}^2$, $\sigma_2^L=-0.01\,\mathrm{e/nm}^2$, $\kappa_1=0.1\,\mathrm{nm}^{-1}$, $\kappa_2=0.03\,\mathrm{nm}^{-1}$, 
$\varepsilon_{r,1}=80$, $\varepsilon_{r,2}=2$, $\beta e\Psi_D=1$, $\ell_B=55.7\,\mathrm{nm}$, $A_1\approx7600\,\mathrm{nm^2}$, 
$A_2\approx22000\,\mathrm{nm^2}$, and $\ell\approx240\,\mathrm{nm}$. The values of the effective 
areas $A_i=4\left(2\kappa_i^{-1}R-\kappa_i^{-2}\right)$ and of the effective length of the three-phase contact line 
$\ell=\sqrt{A_1}+\sqrt{A_2}$ correspond to rough estimates for spherical particles of radii $R=100\,\mathrm{nm}$.
These estimates are obtained by noting that the interaction between two colloidal spheres trapped at the interface 
is essentially given by the interaction between portions of facing caps of height $\kappa_i^{-1}$ inside medium $i$,
so that, by ignoring irrelevant factors of order unity, $A_i$ corresponds to the cap surface area in medium $i$ and
$\ell$ to the averaged length of the intersections between the caps and the interface.
Note that the effective area $A_2$ is larger
than $A_1$ because the inverse Debye length $\kappa_2$ in medium ``2'' is smaller than $\kappa_1$ in medium ``1''. 
From Fig.~\ref{Fig5} one 
infers that the interaction within the superposition calculation is attractive everywhere with a deep minimum ($\approx-155\,k_BT$) at 
$\kappa_1L=0$. Therefore, according to the superposition calculations, the two particles will stick to each other. However, 
the exact calculations predict a strong repulsive interaction between the particles with a shallow minimum ($\approx-1\,k_BT$) at 
separation $\kappa_1L\approx13$ which cannot lead to aggregation of the particles. Thus, depending upon the 
parameters, the results within the two calculations can lead to completely different qualitative behavior. It is important to 
understand that the surface contributions to the total interaction energy $\beta\Delta\widetilde\Omega(L)$ scale with the square of the radii 
of the particles whereas the line part scales linearly with the radii. Consequently, for bigger particles, the surfaces parts 
dominate over the line part. However, for relatively small particles, the line part can easily become comparable to the surface 
parts to alter both the depth and the position of the minimum of the total interaction energy. For example, if one considers a system
with $\sigma_1^0=0.02\,\mathrm{e/nm}^2$, $\sigma_1^L=-0.03\,\mathrm{e/nm}^2$, $\sigma_2^0=-0.0004\,\mathrm{e/nm}^2$, 
$\sigma_2^L=0.0002\,\mathrm{e/nm}^2$, $\kappa_1=0.1\,\mathrm{nm}^{-1}$, $\kappa_2=0.01\,\mathrm{nm}^{-1}$, $\varepsilon_{r,1}=80$, 
$\varepsilon_{r,2}=2$, $\beta e\Psi_D=1$, $\ell_B=55.7\,\mathrm{nm}$, and $R=150\,\mathrm{nm}$, the minimum of the interaction energy 
is around $\kappa_1L\approx6.3$ with a depth of $\approx-57\,k_BT$ when disregarding the line interaction whereas it becomes significantly 
deeper ($\approx-99\,k_BT$) and shifts to $\kappa_1L\approx1.2$ when including the line contribution. Please note that the surface 
interaction in the more polar medium decays faster than the one in the less polar medium and the line interaction. As a result, 
at relatively large separations, the line part competes with the surface interaction in the less polar phase which is weak due 
to small surface charges in the oil phase.

Although the failure of the superposition approximation is not unexpected at short separations, we would like 
to emphasize that the main purpose of this study is to provide exact expressions for the interaction energies valid at short 
separations between particles and to asses the error of the superposition approximation in this limit. Certainly the 
flat wall assumption fails at large separations between spherical particles, but the factor of ``2'' discrepancies in 
this limit for all the three interaction contributions are indeed surprising: For two flat surfaces one usually
assumes the superposition approximation to work well at large separations as the electrostatic potential is screened in both the 
fluid phases, and the influence of each surface on the other is expected to vanish exponentially with increasing separation distance.
The presence of the fluid interface or the heterogeneities of charge densities on each surfaces cannot be the reason for the 
difference since we obtain the factor of ``2'' mismatch for the surface interaction energies which are independent of the 
interface or the surface charge heterogeneity. An open question for future work is the influence of finite particle radii 
on the quantitative mismatch.

\subsubsection{Wall-fluid(s) and fluid-fluid interactions}

Here we discuss the remaining $L$-independent energy contributions to Eq.~(\ref{eq:3}) which, apart from the bulk 
contributions $\Omega_{b,i}$, stem from the wall-fluid(s) interactions or from 
the fluid-fluid interactions. The bulk contribution $\Omega_{b,i}$ is independent of the electrostatic potential inside 
the system and therefore, are the same $\left(-2I_i/\beta\right)$ within both the exact and the superposition 
calculations. Surface tensions acting between the walls placed at $z=0$ or $z=L$ and medium ``1'' are given by 
$\gamma_1^0=\frac{\left(\sigma_1^0\right)^2}{2\kappa_1\varepsilon_1}$ or 
$\gamma_1^L=\frac{\left(\sigma_1^L\right)^2}{2\kappa_1\varepsilon_1}$, respectively, within both the calculations. 
However, for medium ``2'', the surface tensions acting between the wall at $z=0$ and the fluid medium ``2'' within 
the \textit{e}xact calculations is given by $\gamma_2^{0,e}=\frac{\left(\sigma_2^0\right)^2}{2\kappa_2\varepsilon_2}
+\sigma_2^0\Psi_D$, whereas within the \textit{s}uperposition approximation it is given by 
$\gamma_2^{0,s}=\frac{\left(\sigma_2^0\right)^2}{2\kappa_2\varepsilon_2}+\frac{3}{2}\sigma_2^0\Psi_D$. These expressions 
remain the same for $\gamma_2^L$ within the two calculations albeit with $\sigma_2^L$ in place of $\sigma_2^0$. The 
interfacial tension $\gamma_{1,2}$ acting between the two fluids within the two  calculation schemes also differ from each 
other. Whereas the \textit{e}xact calculations give $\gamma_{1,2}^e=-\frac{\kappa_1\kappa_2\varepsilon_1\varepsilon_2\Psi_D^2}
{2\left(\kappa_1\varepsilon_1+\kappa_2\varepsilon_2\right)}$, \textit{s}uperposition calculations lead to an exactly doubled value: 
$\gamma_{1,2}^s=-\frac{\kappa_1\kappa_2\varepsilon_1\varepsilon_2\Psi_D^2}{\kappa_1\varepsilon_1+\kappa_2\varepsilon_2}$. 
The line tension acting at the three-phase contact line at $z=0$ due to an interaction of the wall with the two fluids is 
given within the \textit{e}xact calculations ($\tau^{0,e}$) and the \textit{s}uperposition approximation ($\tau^{0,s}$) by:

\begin{widetext}
 \begin{align}
  \tau^{0,e}=\frac{\kappa_2\varepsilon_2\Psi_D}{\kappa_1\varepsilon_1+\kappa_2\varepsilon_2}
             \left(\frac{\sigma_1^0}{\kappa_1}-\frac{\kappa_1\varepsilon_1}{\kappa_2\varepsilon_2}\frac{\sigma_2^0}{\kappa_2}\right)
             +\frac{1}{\pi}\int\limits_0^{\infty}\frac{dq}{p_1(q)\varepsilon_1+p_2(q)\varepsilon_2}
             \left(\frac{2\sigma_1^0\sigma_2^0}{p_1(q)p_2(q)}
             -\frac{p_2(q)\varepsilon_2}{p_1(q)\varepsilon_1}\left(\frac{\sigma_1^0}{p_1(q)}\right)^2
             -\frac{p_1(q)\varepsilon_1}{p_2(q)\varepsilon_2}\left(\frac{\sigma_2^0}{p_2(q)}\right)^2\right),
  \label{eq:11}
 \end{align}
and
 \begin{align}
  \tau^{0,s}=\frac{3}{2}\frac{\kappa_2\varepsilon_2\Psi_D}{\kappa_1\varepsilon_1+\kappa_2\varepsilon_2}
             \left(\frac{\sigma_1^0}{\kappa_1}-\frac{\kappa_1\varepsilon_1}{\kappa_2\varepsilon_2}\frac{\sigma_2^0}{\kappa_2}\right)
             +\frac{1}{\pi}\int\limits_0^{\infty}\frac{dq}{p_1(q)\varepsilon_1+p_2(q)\varepsilon_2}
             \left(\frac{2\sigma_1^0\sigma_2^0}{p_1(q)p_2(q)}
             -\frac{p_2(q)\varepsilon_2}{p_1(q)\varepsilon_1}\left(\frac{\sigma_1^0}{p_1(q)}\right)^2
             -\frac{p_1(q)\varepsilon_1}{p_2(q)\varepsilon_2}\left(\frac{\sigma_2^0}{p_2(q)}\right)^2\right),
  \label{eq:12}
 \end{align}
\end{widetext}
respectively. Corresponding expressions for $\tau^{L,e}$ and $\tau^{L,s}$ can be obtained by replacing 
$\sigma_i^0$ with $\sigma_i^L$ in the two expressions. Although the integral term is the same in both expressions, 
clearly $\tau^{0,e}\neq\tau^{0,s}$ due to an additional prefactor of $\frac{3}{2}$ present in the first term of Eq.~(\ref{eq:12}). 
This is related to the fact that the exact electrostatic potential $\Psi_i^e(x,z)$ in Eq.~(\ref{eq:4}) reduces 
to the single wall potential $\Psi_i^{\text{sin}}(x,z)$ in Eq.~(\ref{eq:5}) in the limit $L\rightarrow\infty$ 
with $\sigma_2^0=\sigma_2^L=0$, whereas the superposition potential $\Psi_i^s(x,z)$ in Eq.~(\ref{eq:6}) does not. 

\section{Conclusions}

In summary, by using a classical DFT approach, we have analyzed the electrostatic interaction between two unequally 
charged parallel walls in contact with two immiscible liquids. Within the framework of a linearized PB mean-field 
theory, we have derived analytical expressions for the electrostatic potential distribution inside the system both 
within and by going beyond the linear superposition approximation. Then these potentials are used to calculate the 
surface and line interaction energy densities between the walls. As an important finding we see that both the 
surface and the line interaction can vary monotonically and non-monotonically and in most cases, the superposition 
approximation fails to predict the correct variations at short separations. Moreover, contrary to common assumption, 
the superposition approximation can be quantitatively as well as qualitatively incorrect even at large separations.
Analytical expressions are also provided for other interaction parameters of the system, i.e., the surface tensions, 
the line tensions and the fluid-fluid interfacial tension. It turns out that all these constant (independent of 
the separation distance between the walls) interaction parameters also differ more or less within the two calculation 
schemes. The system under consideration is expected to mimic the interaction between two dissimilar colloids trapped 
at an electrolyte interface with a surface to surface separation distance small compared to the radii of the particles. 
Not only that, our general study can also be applied to other situations like the interaction between two Janus colloids 
in the bulk or at an interface or a single Janus particle approaching a solid surface \cite{Ras17}.


\end{document}